# Lower Bounds on Q for Finite Size Antennas of Arbitrary Shape

Oleksiy S. Kim

***Abstract*—The problem of the lower bound on the radiation $Q$ for an arbitrarily shaped finite size antenna of non-zero volume is formulated in terms of equivalent electric and magnetic currents densities distributed on a closed surface coinciding with antenna exterior surface. When these equivalent currents radiate in free space, the magnetic current augments the electric current, so that the fields interior to the surface vanish. In contrast to approaches based solely on electric currents, the proposed technique ensures no stored energy interior to the antenna exterior surface, and thus, allows the fundamental lower bound on $Q$ to be determined. To facilitate the computation of the bound, new expressions for the stored energy, radiated power, and $Q$ of coupled electric and magnetic source currents in free space are derived.**

## I. Introduction

PHYSICAL limitations on the bandwidth of electrically small antennas are normally established in terms of lower bounds on the antenna radiation quality factor $Q$. Among the shapes of finite size, the problem has been solved in closed form only for a sphere [1]–[3] and an infinitely long cylinder [4]. In the limit of vanishingly small antennas, the range of closed-form bounds is wider and includes truncated cylinders [5], circular disks, needles, and toroidal rings [6] as well as various spheroids [7]. For other shapes, the lower bound on $Q$ has to be found numerically by solving either scattering or radiation problem.

The scattering approach by Gustafsson *et al.* [8] involves a free parameter that needs to be set empirically, whereas the radiation approach [9]–[11] is more robust and does not require any calibration. On the other hand, the radiation approach that is based on the expressions for the $Q$ of an electric source current radiating in free space [12] is generally applicable only for antennas of zero volume, such as thin-sheet or thin-wire antennas. An attempt to determine $Q$ for an antenna shape of finite volume using solely equivalent electric currents on its surface will not result in a fundamental lower bound, because it will include the energy stored in the shape's volume. The bound will be valid for air-core antennas, for example, spherical wire antennas [13], [14], but not in general. Indeed, spherical dipole antennas with magnetic cores can exhibit $Q$'s not just below the air-core bounds [15], [16], but very close to the Chu lower bound [17]–[19] and even to the fundamental lower bound [20], which no passive linear time-invariant antenna can overcome. The $Q$ for a cylindrical dipole antenna was also shown able to go below its air-core bound [17]. This means that to find the absolute lower bound for a given shape the interior stored energy must be excluded.

This paper presents an approach to determining the lower bound on $Q$ for an arbitrary finite size antenna shape based on equivalent electric and magnetic current densities on the antenna exterior surface, whose respective radiation mutually cancel inside this surface. The resulting $Q$ is then the true lower bound for a given shape. In [7] (with corrections in [21]), this approach was applied to vanishingly small antennas; here, it is extended to antennas of finite size. To implement the approach, two problems have been solved:

1) Closed-form expressions for the $Q$ of coupled electric and magnetic currents in free space have been derived without any approximation (Section II).
2) A procedure for computing the magnetic current density given the electric current density on the antenna surface, such that the fields interior to the surface vanish, has been established (Section III).

The main theoretical results are summarized in Tables I and II that provide a complete set of expressions necessary to evaluate the stored electric and magnetic energies as well as the radiated power, and thus $Q$, for any combination of electric and magnetic source currents in free space.

Besides solving the problem of the lower bound on $Q$, the presented expressions and methods allow the $Q$ of metal-dielectric antennas to be computed using equivalent electric and magnetic current densities on their surfaces.

## II. Stored Energy and Radiation Q for Electric and Magnetic Source Currents

The radiation $Q$ defined for a lossless antenna as

$$Q = 2\omega \frac{\max\left(W^{\mathrm{e}}, W^{\mathrm{m}}\right)}{P^{\mathrm{rad}}} \tag{1}$$

where $\omega$ is the angular frequency, requires the stored electric $W^{\mathrm{e}}$ and magnetic $W^{\mathrm{m}}$ energies as well as the radiated power $P^{\mathrm{rad}}$ to be determined first.

### A. Stored Energy

Following the procedure of [3], [22], [23], we will derive the stored energy associated with electric and magnetic currents distributed in volume $V$ by integrating over the entire space $V_\infty$ the difference between the total energy density and the energy density of the propagating field in free space. First, we

The author is with the Department of Electrical Engineering, Electromagnetic Systems, Technical University of Denmark, DK-2800 Kgs. Lyngby, Denmark (e-mail: osk@elektro.dtu.dk)

will do it for the stored electric energy. The stored magnetic energy is then easily obtained by the duality transformation.

Electric $\boldsymbol{E}(\boldsymbol{r})$ and magnetic $\boldsymbol{H}(\boldsymbol{r})$ fields radiated by electric $\boldsymbol{J}(\boldsymbol{r})$ and magnetic $\boldsymbol{M}(\boldsymbol{r})$ current densities in free space are given by

$$\boldsymbol{E}(\boldsymbol{r}) = -j\omega \boldsymbol{A}(\boldsymbol{r}) - \nabla\Phi(\boldsymbol{r}) - \frac{1}{\varepsilon_0}\nabla\times\boldsymbol{F}(\boldsymbol{r}) \tag{2}$$

$$\boldsymbol{H}(\boldsymbol{r}) = -j\omega \boldsymbol{F}(\boldsymbol{r}) - \nabla\Phi_m(\boldsymbol{r}) + \frac{1}{\mu_0}\nabla\times\boldsymbol{A}(\boldsymbol{r}) \tag{3}$$

where $\boldsymbol{r}$ denotes the position vector. The vector and scalar potentials are

$$\boldsymbol{A}(\boldsymbol{r}) = \mu_0 \int\limits_V \boldsymbol{J}(\boldsymbol{r}')G(\boldsymbol{r},\boldsymbol{r}')dV' \tag{4a}$$

$$\boldsymbol{F}(\boldsymbol{r}) = \varepsilon_0 \int\limits_V \boldsymbol{M}(\boldsymbol{r}')G(\boldsymbol{r},\boldsymbol{r}')dV' \tag{4b}$$

$$\Phi(\boldsymbol{r}) = \frac{j}{\omega\varepsilon_0} \int\limits_V \nabla'\cdot\boldsymbol{J}(\boldsymbol{r}')G(\boldsymbol{r},\boldsymbol{r}')dV' \tag{4c}$$

$$\Phi_m(\boldsymbol{r}) = \frac{j}{\omega\mu_0} \int\limits_V \nabla'\cdot\boldsymbol{M}(\boldsymbol{r}')G(\boldsymbol{r},\boldsymbol{r}')dV' \tag{4d}$$

where $\varepsilon_0$ and $\mu_0$ are permittivity and permeability of free space, respectively, and $G(\boldsymbol{r},\boldsymbol{r}') = \exp(-jk|\boldsymbol{r}-\boldsymbol{r}'|)/(4\pi|\boldsymbol{r}-\boldsymbol{r}'|)$ is the scalar Green's function of free space with $k = \omega\sqrt{\varepsilon_0\mu_0}$ being the free space propagation constant; the free-space impedance is denoted by $\eta_0 = \sqrt{\mu_0/\varepsilon_0}$. The time factor $\exp(j\omega t)$, where $t$ is the time variable, is assumed and suppressed.

The electric far-field is

$$\boldsymbol{E}_\infty(\hat{\boldsymbol{r}}) = -j\omega\boldsymbol{A}_\infty(\hat{\boldsymbol{r}}) + \hat{\boldsymbol{r}}jk\Phi_\infty(\hat{\boldsymbol{r}}) + \frac{jk}{\varepsilon_0}\hat{\boldsymbol{r}}\times\boldsymbol{F}_\infty(\hat{\boldsymbol{r}}) \tag{5}$$

where $\hat{\boldsymbol{r}} = \boldsymbol{r}/|\boldsymbol{r}|$ is a unit vector pointing towards an observation point in space and the 'far-field' potentials $\boldsymbol{A}_\infty(\hat{\boldsymbol{r}})$, $\boldsymbol{F}_\infty(\hat{\boldsymbol{r}})$, and $\Phi_\infty(\hat{\boldsymbol{r}})$ are expressed as in (4) with the Green's function $G(\boldsymbol{r},\boldsymbol{r}')$ replaced by its far-field counterpart $G_\infty(\hat{\boldsymbol{r}},\boldsymbol{r}') = \exp(jk\boldsymbol{r}'\cdot\hat{\boldsymbol{r}} - jk|\boldsymbol{r}|)/(4\pi|\boldsymbol{r}|)$.

The total electric energy density $w^{\mathrm{e}}$ can now be found using (2) as

$$\frac{4w^{\mathrm{e}}}{\varepsilon_0} = |\boldsymbol{E}|^2 = \boldsymbol{E}\cdot\boldsymbol{E}^* = |j\omega\boldsymbol{A}+\nabla\Phi|^2 + \frac{1}{\varepsilon_0^2}|\nabla\times\boldsymbol{F}|^2 - \frac{2}{\varepsilon_0}\mathrm{Re}\{j\omega\boldsymbol{A}^*\cdot[\nabla\times\boldsymbol{F}]\} + \frac{2}{\varepsilon_0}\mathrm{Re}\{\nabla\Phi^*\cdot[\nabla\times\boldsymbol{F}]\} \tag{6}$$

whereas the energy density of the propagating field $w^{\mathrm{e}}_\infty$ reads as

$$\frac{4w^{\mathrm{e}}_\infty}{\varepsilon_0} = |\boldsymbol{E}_\infty|^2 = \omega^2|\boldsymbol{A}_\infty|^2 - k^2|\Phi_\infty| + \frac{k^2}{\varepsilon_0^2}|\hat{\boldsymbol{r}}\times\boldsymbol{F}_\infty|^2 - \frac{2\omega k}{\varepsilon_0}\mathrm{Re}\{\boldsymbol{A}^*_\infty\cdot[\hat{\boldsymbol{r}}\times\boldsymbol{F}_\infty]\} \tag{7}$$

where asterisk '*' denotes complex conjugate. Expression (7) was written using (5) and the fact that the far-field $\boldsymbol{E}_\infty(\hat{\boldsymbol{r}})$ and the direction vector $\hat{\boldsymbol{r}}$ are orthogonal.

Subtracting (7) from (6) and integrating the result over the entire space $V_\infty$, we can express the total stored electric energy $W^{\mathrm{e}}_{\mathrm{tot}}$ as a sum of coordinate-independent $W^{\mathrm{e}}$ and coordinate-dependent $W^{\mathrm{e}}_{\mathrm{cd}}$ terms. Each of these is a sum of contributions due to electric, magnetic, and coupled electric and magnetic currents.

$$\begin{aligned} W^{\mathrm{e}}_{\mathrm{tot}} &= \frac{\varepsilon_0}{4}\int\limits_{V_\infty} |\boldsymbol{E}|^2 - |\boldsymbol{E}_\infty|^2 \mathrm{dV} \\ &= \frac{\varepsilon_0}{4}\int\limits_{V_\infty} |j\omega\boldsymbol{A}+\nabla\Phi|^2 - \omega^2|\boldsymbol{A}_\infty|^2 + k^2|\Phi_\infty|\mathrm{dV} \\ &+ \frac{1}{4\varepsilon_0}\int\limits_{V_\infty} |\nabla\times\boldsymbol{F}|^2 - k^2|\hat{\boldsymbol{r}}\times\boldsymbol{F}_\infty|^2\mathrm{dV} \\ &- \frac{\omega}{2}\int\limits_{V_\infty} \mathrm{Re}\{j\boldsymbol{A}^*\cdot[\nabla\times\boldsymbol{F}] - k\boldsymbol{A}^*_\infty\cdot[\hat{\boldsymbol{r}}\times\boldsymbol{F}_\infty]\}\mathrm{dV} \\ &= W^{\mathrm{e}} + W^{\mathrm{e}}_{\mathrm{cd}} = W^{\mathrm{e}}_{\mathrm{J}} + W^{\mathrm{e}}_{\mathrm{M}} + W^{\mathrm{e}}_{\mathrm{JM}} + W^{\mathrm{e}}_{\mathrm{cd}}. \end{aligned} \tag{8}$$

Note that the integration of the last term in (6) gives zero (see Appendix A for details).

An expression for $W^{\mathrm{e}}_{\mathrm{J}}$ was rigorously derived in [12], while an expression for $W^{\mathrm{e}}_{\mathrm{M}}$ can be obtained in a straightforward manner from the expression for the stored magnetic energy associated with electric current $W^{\mathrm{m}}_{\mathrm{J}}$, also derived in [12], by applying the duality transformation. The contribution to the total stored electric energy due to coupled electric and magnetic currents is derived here.

Substituting (4) into (8), we find

$$\begin{aligned} W^{\mathrm{e}}_{\mathrm{JM,tot}} &= W^{\mathrm{e}}_{\mathrm{JM}} + W^{\mathrm{e}}_{\mathrm{JM,cd}} \\ &= \frac{k^2}{2\omega}\int\limits_{V_\infty} \mathrm{Re}\Bigg\{\int\limits_V j\boldsymbol{J}^*_2G^*_2\mathrm{dV}_2\cdot\int\limits_V[\boldsymbol{M}_1\times\nabla G_1]\mathrm{dV}_1 \\ &+ k\int\limits_V \boldsymbol{J}^*_2G^*_{2\infty}\mathrm{dV}_2\cdot\int\limits_V[\hat{\boldsymbol{r}}\times\boldsymbol{M}_1]G^*_{1\infty}\mathrm{dV}_1\Bigg\}\mathrm{dV} \end{aligned} \tag{9a}$$

$$\begin{aligned} &= \frac{k^2}{2\omega}\int\limits_V\int\limits_V \mathrm{Re}\Bigg\{[\boldsymbol{J}^*_2\times\boldsymbol{M}_1]\cdot\int\limits_{V_\infty} j(\nabla G_1)G^*_2 \\ &- \hat{\boldsymbol{r}}k\frac{e^{jk(\boldsymbol{r}_1-\boldsymbol{r}_2)\cdot\hat{\boldsymbol{r}}}}{16\pi^2|\boldsymbol{r}|^2}\mathrm{dV}\Bigg\}\mathrm{dV}_1\mathrm{dV}_2 \end{aligned} \tag{9b}$$

where subscripts 1 and 2 refer to the first and second integration over the same volume $V$ containing the source currents $\boldsymbol{J}$ and $\boldsymbol{M}$. The inner integral in (9b) integrates the difference between the total and propagating energy densities at every point in space. The latter density depends on the choice of the origin of the coordinate system giving rise to the coordinate-dependence in the expression for the total stored energy. By evaluating the integral in closed form (Appendix B), we can separate the coordinate-independent and coordinate-dependent terms[1].

In the following we will consider only the coordinate-independent part of the total stored energy. In practice, this is justified by the fact the coordinate-dependent contribution to $Q$ can be kept below $ka$ by placing the origin of the coordinate system within the minimum sphere of radius $a$ circumscribing an antenna [24]. The resulting coordinate-dependent contribution thus only becomes significant at rather large $ka$, where the inverse relationship between the $Q$ and the frequency bandwidth breaks down anyway.

[1]A similar coordinate-dependent term appearing in $W^{\mathrm{e}}_{\mathrm{J,tot}}$ was treated in [23]

A new expression for the coordinate-independent part of the stored electric energy associated with coupled electric and magnetic currents reads as (Appendix B)

$$W_{\mathrm{JM}}^{\mathrm{e}} = \frac{1}{4\omega}\int_V\int_V \mathrm{Re}[\boldsymbol{M}_1 \times \boldsymbol{J}_2^*]\cdot \mathrm{Im}\{\nabla_1 G_{12}\} + k^2 \mathrm{Im}[\boldsymbol{M}_1 \times \boldsymbol{J}_2^*]\cdot \boldsymbol{r}_{12}\,\mathrm{Re}\{G_{12}\}\mathrm{dV}_1\mathrm{dV}_2 \tag{10a}$$

$$= \frac{1}{16\pi\omega}\int_V\int_V \mathrm{Re}[\boldsymbol{M}_1 \times \boldsymbol{J}_2^*]\cdot \boldsymbol{r}_{12}\left(\frac{\sin(k|\boldsymbol{r}_{12}|)}{|\boldsymbol{r}_{12}|^3} - \frac{k\cos(k|\boldsymbol{r}_{12}|)}{|\boldsymbol{r}_{12}|^2}\right) + \mathrm{Im}[\boldsymbol{M}_1 \times \boldsymbol{J}_2^*]\cdot \boldsymbol{r}_{12}\frac{k^2\cos(k|\boldsymbol{r}_{12}|)}{|\boldsymbol{r}_{12}|}\mathrm{dV}_1\mathrm{dV}_2 \tag{10b}$$

where $G_{12} = G(\boldsymbol{r}_1, \boldsymbol{r}_2)$ and $\boldsymbol{r}_{12} = \boldsymbol{r}_1 - \boldsymbol{r}_2$.

Thus, the stored electric energy $W^{\mathrm{e}}$ can be evaluated using (8) together with (10) and [12, equations (63), (64)].

The corresponding expressions for the stored magnetic energy $W_{\mathrm{J}}^{\mathrm{m}}$, $W_{\mathrm{M}}^{\mathrm{m}}$, and $W_{\mathrm{JM}}^{\mathrm{m}}$ are obtained from the expressions for $W_{\mathrm{M}}^{\mathrm{e}}$, $W_{\mathrm{J}}^{\mathrm{e}}$, and $W_{\mathrm{JM}}^{\mathrm{e}}$, respectively, by the duality transformation. The results for both $W^{\mathrm{e}}$ and $W^{\mathrm{m}}$ are summarized in Table I providing an overview of the expressions necessary to evaluate the stored energy for any combination of electric and magnetic source currents in free space.

### B. Poynting's Theorem and Radiated Power

Poynting's theorem relates the supplied power to the radiated power $P^{\mathrm{rad}}$ and the stored energy as

$$-\frac{1}{2}\int_V \boldsymbol{E}\cdot\boldsymbol{J}^* + \boldsymbol{M}\cdot\boldsymbol{H}^*\,\mathrm{dV} = P^{\mathrm{rad}} + j2\omega(W^{\mathrm{m}} - W^{\mathrm{e}}). \tag{11}$$

It can be easily verified by substituting the energies from Table I and fields from (2) and (3) into (11) that the following identity holds

$$-\frac{1}{2}\int_V \mathrm{Im}\left\{\boldsymbol{E}\cdot\boldsymbol{J}^* + \boldsymbol{M}\cdot\boldsymbol{H}^*\right\}\mathrm{dV} = 2\omega(W^{\mathrm{m}} - W^{\mathrm{e}}) \tag{12}$$

thus, confirming the validity of the expression derived in Section II-A.

The real part of (11) constitutes the radiated power, which can be represented as a sum of three contributions, similar to the stored energy (8)

$$P^{\mathrm{rad}} = -\frac{1}{2}\int_V \mathrm{Re}\left\{\boldsymbol{E}\cdot\boldsymbol{J}^* + \boldsymbol{M}\cdot\boldsymbol{H}^*\right\}\mathrm{dV} = P_{\mathrm{J}}^{\mathrm{rad}} + P_{\mathrm{M}}^{\mathrm{rad}} + P_{\mathrm{JM}}^{\mathrm{rad}}. \tag{13}$$

An expression for the radiated power $P_{\mathrm{J}}^{\mathrm{rad}}$ due to the electric current $\boldsymbol{J}$ was derived in [12], from which an expression for $P_{\mathrm{M}}^{\mathrm{rad}}$, denoting the power radiated by the magnetic current $\boldsymbol{M}$, can again be obtained via the duality transformation.

The radiated power $P_{\mathrm{JM}}^{\mathrm{rad}}$ due to the coupled electric and magnetic currents is found by inserting (2) and (3) into (13) and isolating the terms containing cross products of $\boldsymbol{J}$ and $\boldsymbol{M}$ as

$$P_{\mathrm{JM}}^{\mathrm{rad}} = \int_V\int_V \mathrm{Im}[\boldsymbol{M}_1 \times \boldsymbol{J}_2^*]\cdot\mathrm{Im}\{\nabla_1 G_{12}\}\mathrm{dV}_1\mathrm{dV}_2 = \frac{1}{4\pi}\int_V\int_V \mathrm{Im}[\boldsymbol{M}_1 \times \boldsymbol{J}_2^*]\cdot\boldsymbol{r}_{12}\left(\frac{\sin(k|\boldsymbol{r}_{12}|)}{|\boldsymbol{r}_{12}|^3} - \frac{k\cos(k|\boldsymbol{r}_{12}|)}{|\boldsymbol{r}_{12}|^2}\right)\mathrm{dV}_1\mathrm{dV}_2. \tag{14}$$

Contrary to the stored energy in (10), this power term does not vanish with the size of the antenna [25].

All the expressions for the radiated power terms $P_{\mathrm{J}}^{\mathrm{rad}}$, $P_{\mathrm{M}}^{\mathrm{rad}}$, and $P_{\mathrm{JM}}^{\mathrm{rad}}$ are provided in Table I.

### C. Integral Operator and Matrix Representation

The expressions in Table I may look cumbersome, but their implementation into an existing method of moments code is straightforward. It is even more convenient when the expressions are recast into an operator/matrix notation, as in Table II. Operators $\mathcal{L}_a$, $\mathcal{L}_f$, and $\mathcal{K}$ arise in electric field and magnetic field integral equation (EFIE and MFIE, respectively) formulations [26], while operators $\mathcal{L}_c$ and $\mathcal{K}_c$ are simple modifications of the first three without $1/r$ singularities.

The expressions '$\langle *, * \rangle$' in Table II can also be interpreted as vector-matrix-vector products, where the vectors contain coefficients of the expansion of the respective currents into basis functions and the matrices are discretized (with Galerkin's method) representations of the respective integral operators[2]. In a combined-field integral equation code (for example, based on PMCHWT [27] or Muller's [28] formulation for homogeneous dielectric objects), the matrices corresponding to operators $\mathcal{L}_a$, $\mathcal{L}_f$, and $\mathcal{K}$ are naturally computed; computing the matrices for operators $\mathcal{L}_c$ and $\mathcal{K}_c$, thus, imposes almost no overhead.

## III. Electric and Magnetic Currents for Zero Interior Fields

Now the task is to find electric and magnetic currents on the surface of a volumetric shape, for which the lower bound on $Q$ is sought, such that the fields inside this shape vanish. This is accomplished by enforcing a boundary condition for electric and magnetic surface current densities, $\boldsymbol{J}_s$ and $\boldsymbol{M}_s$, at the exterior surface $S$ of the object as [29], [30]

$$\left[\omega\mu_0(\mathcal{L}_a - \mathcal{L}_f)\boldsymbol{J}_s - \left(\mathcal{K} - \frac{1}{2}\mathcal{N}\right)\boldsymbol{M}_s\right]_{\mathrm{tan}} = 0 \tag{15a}$$

$$\left[\left(\mathcal{K} - \frac{1}{2}\mathcal{N}\right)\boldsymbol{J}_s + \omega\varepsilon_0(\mathcal{L}_a - \mathcal{L}_f)\boldsymbol{M}_s\right]_{\mathrm{tan}} = 0 \tag{15b}$$

where $\mathcal{N}\boldsymbol{X} = \hat{\boldsymbol{n}} \times \boldsymbol{X}$ with $\hat{\boldsymbol{n}}$ being an outward normal to the surface $S$; 'tan' denotes the components tangential to $S$. Expanding $\boldsymbol{J}_s$ and $\boldsymbol{M}_s$ in terms of a set of basis functions and

[2] Matrices arising from $\mathcal{L}_a$, $\mathcal{L}_f$, and $\mathcal{L}_c$ are generally different for electric and magnetic source currents

TABLE I

RADIAITON $Q$, STORED ENERGY, AND RADIATED POWER FOR ELECTRIC AND MAGNETIC SOURCE CURRENTS IN FREE SPACE

Electric source current $\boldsymbol{J}(\boldsymbol{r})$

$$Q_{\mathrm{J}} = \frac{2\omega \max(W_{\mathrm{J}}^{\mathrm{e}}, W_{\mathrm{J}}^{\mathrm{m}})}{P_{\mathrm{J}}^{\mathrm{rad}}}$$

$$W_{\mathrm{J}}^{\mathrm{e}} = \frac{\mu_0}{4} \int_V \int_V \frac{1}{k^2} \nabla_1 \cdot \boldsymbol{J}_1 \nabla_2 \cdot \boldsymbol{J}_2^* \frac{\cos(k|\boldsymbol{r}_{12}|)}{4\pi|\boldsymbol{r}_{12}|} - \frac{1}{2}\left(k \boldsymbol{J}_1 \cdot \boldsymbol{J}_2^* - \frac{1}{k} \nabla_1 \cdot \boldsymbol{J}_1 \nabla_2 \cdot \boldsymbol{J}_2^*\right) \frac{\sin(k|\boldsymbol{r}_{12}|)}{4\pi} \, \mathrm{dV}_1 \mathrm{dV}_2$$

$$W_{\mathrm{J}}^{\mathrm{m}} = \frac{\mu_0}{4} \int_V \int_V \boldsymbol{J}_1 \cdot \boldsymbol{J}_2^* \frac{\cos(k|\boldsymbol{r}_{12}|)}{4\pi|\boldsymbol{r}_{12}|} - \frac{1}{2}\left(k \boldsymbol{J}_1 \cdot \boldsymbol{J}_2^* - \frac{1}{k} \nabla_1 \cdot \boldsymbol{J}_1 \nabla_2 \cdot \boldsymbol{J}_2^*\right) \frac{\sin(k|\boldsymbol{r}_{12}|)}{4\pi} \, \mathrm{dV}_1 \mathrm{dV}_2$$

$$P_{\mathrm{J}}^{\mathrm{rad}} = \frac{\eta_0}{2} \int_V \int_V \left(k \boldsymbol{J}_1 \cdot \boldsymbol{J}_2^* - \frac{1}{k} \nabla_1 \cdot \boldsymbol{J}_1 \nabla_2 \cdot \boldsymbol{J}_2^*\right) \frac{\sin(k|\boldsymbol{r}_{12}|)}{4\pi|\boldsymbol{r}_{12}|} \, \mathrm{dV}_1 \mathrm{dV}_2$$

Magnetic source current $\boldsymbol{M}(\boldsymbol{r})$

$$Q_{\mathrm{M}} = \frac{2\omega \max(W_{\mathrm{M}}^{\mathrm{e}}, W_{\mathrm{M}}^{\mathrm{m}})}{P_{\mathrm{M}}^{\mathrm{rad}}}$$

$$W_{\mathrm{M}}^{\mathrm{e}} = \frac{\varepsilon_0}{4} \int_V \int_V \boldsymbol{M}_1 \cdot \boldsymbol{M}_2^* \frac{\cos(k|\boldsymbol{r}_{12}|)}{4\pi|\boldsymbol{r}_{12}|} - \frac{1}{2}\left(k \boldsymbol{M}_1 \cdot \boldsymbol{M}_2^* - \frac{1}{k} \nabla_1 \cdot \boldsymbol{M}_1 \nabla_2 \cdot \boldsymbol{M}_2^*\right) \frac{\sin(k|\boldsymbol{r}_{12}|)}{4\pi} \, \mathrm{dV}_1 \mathrm{dV}_2$$

$$W_{\mathrm{M}}^{\mathrm{m}} = \frac{\varepsilon_0}{4} \int_V \int_V \frac{1}{k^2} \nabla_1 \cdot \boldsymbol{M}_1 \nabla_2 \cdot \boldsymbol{M}_2^* \frac{\cos(k|\boldsymbol{r}_{12}|)}{4\pi|\boldsymbol{r}_{12}|} - \frac{1}{2}\left(k \boldsymbol{M}_1 \cdot \boldsymbol{M}_2^* - \frac{1}{k} \nabla_1 \cdot \boldsymbol{M}_1 \nabla_2 \cdot \boldsymbol{M}_2^*\right) \frac{\sin(k|\boldsymbol{r}_{12}|)}{4\pi} \, \mathrm{dV}_1 \mathrm{dV}_2$$

$$P_{\mathrm{M}}^{\mathrm{rad}} = \frac{1}{2\eta_0} \int_V \int_V \left(k \boldsymbol{M}_1 \cdot \boldsymbol{M}_2^* - \frac{1}{k} \nabla_1 \cdot \boldsymbol{M}_1 \nabla_2 \cdot \boldsymbol{M}_2^*\right) \frac{\sin(k|\boldsymbol{r}_{12}|)}{4\pi|\boldsymbol{r}_{12}|} \, \mathrm{dV}_1 \mathrm{dV}_2$$

Electric and magnetic source currents, $\boldsymbol{J}(\boldsymbol{r})$ and $\boldsymbol{M}(\boldsymbol{r})$

$$Q = \frac{2\omega \max(W^{\mathrm{e}}, W^{\mathrm{m}})}{P^{\mathrm{rad}}}; \quad W^{\mathrm{e}} = W_{\mathrm{J}}^{\mathrm{e}} + W_{\mathrm{M}}^{\mathrm{e}} + W_{\mathrm{JM}}^{\mathrm{e}}; \; W^{\mathrm{m}} = W_{\mathrm{J}}^{\mathrm{m}} + W_{\mathrm{M}}^{\mathrm{m}} + W_{\mathrm{JM}}^{\mathrm{m}}; \; P^{\mathrm{rad}} = P_{\mathrm{J}}^{\mathrm{rad}} + P_{\mathrm{M}}^{\mathrm{rad}} + P_{\mathrm{JM}}^{\mathrm{rad}}$$

$$W_{\mathrm{JM}}^{\mathrm{e}} = \frac{1}{4\omega} \int_V \int_V \mathrm{Re}[\boldsymbol{M}_1 \times \boldsymbol{J}_2^*] \cdot \frac{\boldsymbol{r}_{12}}{|\boldsymbol{r}_{12}|} \left(\frac{\sin(k|\boldsymbol{r}_{12}|)}{4\pi|\boldsymbol{r}_{12}|^2} - \frac{k\cos(k|\boldsymbol{r}_{12}|)}{4\pi|\boldsymbol{r}_{12}|}\right) + \mathrm{Im}[\boldsymbol{M}_1 \times \boldsymbol{J}_2^*] \cdot \boldsymbol{r}_{12} \frac{k^2 \cos(k|\boldsymbol{r}_{12}|)}{4\pi|\boldsymbol{r}_{12}|} \, \mathrm{dV}_1 \mathrm{dV}_2$$

$$W_{\mathrm{JM}}^{\mathrm{m}} = \frac{1}{4\omega} \int_V \int_V -\mathrm{Re}[\boldsymbol{M}_1 \times \boldsymbol{J}_2^*] \cdot \frac{\boldsymbol{r}_{12}}{|\boldsymbol{r}_{12}|} \left(\frac{\sin(k|\boldsymbol{r}_{12}|)}{4\pi|\boldsymbol{r}_{12}|^2} - \frac{k\cos(k|\boldsymbol{r}_{12}|)}{4\pi|\boldsymbol{r}_{12}|}\right) + \mathrm{Im}[\boldsymbol{M}_1 \times \boldsymbol{J}_2^*] \cdot \boldsymbol{r}_{12} \frac{k^2 \cos(k|\boldsymbol{r}_{12}|)}{4\pi|\boldsymbol{r}_{12}|} \, \mathrm{dV}_1 \mathrm{dV}_2$$

$$P_{\mathrm{JM}}^{\mathrm{rad}} = \int_V \int_V \mathrm{Im}[\boldsymbol{M}_1 \times \boldsymbol{J}_2^*] \cdot \frac{\boldsymbol{r}_{12}}{|\boldsymbol{r}_{12}|} \left(\frac{\sin(k|\boldsymbol{r}_{12}|)}{4\pi|\boldsymbol{r}_{12}|^2} - \frac{k\cos(k|\boldsymbol{r}_{12}|)}{4\pi|\boldsymbol{r}_{12}|}\right) \mathrm{dV}_1 \mathrm{dV}_2$$

applying Galerkin's testing to (15) leads to a system of linear equations, which can be written in matrix-vector form as

$$\begin{bmatrix} \mathbf{Z} & -\mathbf{K}+\mathbf{N} \\ \mathbf{K}-\mathbf{N} & \mathbf{Y} \end{bmatrix} \begin{bmatrix} \mathbf{I} \\ \mathbf{M} \end{bmatrix} = 0 \tag{16}$$

where [**Z**], [**Y**], [**K**], and [**N**] are matrices due to the respective operators in (15); **I** and **M** are vectors containing expansion coefficients for $\boldsymbol{J}_s$ and $\boldsymbol{M}_s$, respectively.

Using the second equation in (16), the coefficients **M**, ensuring zero total magnetic field inside $S$, can be expressed in terms of the **I** coefficients as

$$\mathbf{M} = [\mathbf{Y}]^{-1}[\mathbf{N}-\mathbf{K}]\mathbf{I}\,. \tag{17}$$

Zero magnetic field implies that the electric field also vanishes inside $S$.[3]

## IV. THE LOWER BOUND ON $Q$

The problem of determining the minimum possible $Q$ for an arbitrary volumetric shape is equivalent to identifying an optimal distribution of electric and magnetic currents on its surface. Once the currents are known, the lower bound on $Q$ for the given shape can readily be computed using expressions derived in Section II (Table I or II).

A straightforward way to find the optimal currents is to solve an optimization problem for minimum $Q(\boldsymbol{J},\boldsymbol{M})$ using

[3]Spurious fields may arise at frequencies close to interior cavity resonances. However, these resonances occur beyond electrically small regime, and thus, are not of concern.

TABLE II

STORED ENERGY AND RADIATED POWER FOR ELECTRIC AND MAGNETIC SOURCE CURRENTS IN FREE SPACE. OPERATOR/MATRIX REPRESENTATION

| Electric source current $\boldsymbol{J}(\boldsymbol{r})$ | Magnetic source current $\boldsymbol{M}(\boldsymbol{r})$ | Electric and magnetic source currents |
|---|---|---|
| $W_{\mathrm{J}}^{\mathrm{e}} = \frac{\mu_0}{4} \mathrm{Im} \langle \boldsymbol{J}, (\mathcal{L}_f + \mathcal{L}_c) \boldsymbol{J} \rangle$ | $W_{\mathrm{M}}^{\mathrm{e}} = \frac{\varepsilon_0}{4} \mathrm{Im} \langle \boldsymbol{M}, (\mathcal{L}_a + \mathcal{L}_c) \boldsymbol{M} \rangle$ | $W_{\mathrm{JM}}^{\mathrm{e}} = \frac{1}{8\omega} \left( \mathrm{Im} \{ \langle \boldsymbol{J}, \mathcal{K}\boldsymbol{M} \rangle + \langle \boldsymbol{M}, \mathcal{K}\boldsymbol{J} \rangle \} + \mathrm{Im} \{ \langle \boldsymbol{J}, \mathcal{K}_c\boldsymbol{M} \rangle - \langle \boldsymbol{M}, \mathcal{K}_c\boldsymbol{J} \rangle \} \right)$ |
| $W_{\mathrm{J}}^{\mathrm{m}} = \frac{\mu_0}{4} \mathrm{Im} \langle \boldsymbol{J}, (\mathcal{L}_a + \mathcal{L}_c) \boldsymbol{J} \rangle$ | $W_{\mathrm{M}}^{\mathrm{m}} = \frac{\varepsilon_0}{4} \mathrm{Im} \langle \boldsymbol{M}, (\mathcal{L}_f + \mathcal{L}_c) \boldsymbol{M} \rangle$ | $W_{\mathrm{JM}}^{\mathrm{m}} = \frac{1}{8\omega} \left( -\mathrm{Im} \{ \langle \boldsymbol{J}, \mathcal{K}\boldsymbol{M} \rangle + \langle \boldsymbol{M}, \mathcal{K}\boldsymbol{J} \rangle \} + \mathrm{Im} \{ \langle \boldsymbol{J}, \mathcal{K}_c\boldsymbol{M} \rangle - \langle \boldsymbol{M}, \mathcal{K}_c\boldsymbol{J} \rangle \} \right)$ |
| $P_{\mathrm{J}}^{\mathrm{rad}} = \frac{\eta_0 k}{2} \mathrm{Re} \langle \boldsymbol{J}, (\mathcal{L}_a - \mathcal{L}_f) \boldsymbol{J} \rangle$ | $P_{\mathrm{M}}^{\mathrm{rad}} = \frac{k}{2\eta_0} \mathrm{Re} \langle \boldsymbol{M}, (\mathcal{L}_a - \mathcal{L}_f) \boldsymbol{M} \rangle$ | $P_{\mathrm{JM}}^{\mathrm{rad}} = \frac{1}{2} \mathrm{Re} \{ \langle \boldsymbol{M}, \mathcal{K}\boldsymbol{J} \rangle - \langle \boldsymbol{J}, \mathcal{K}\boldsymbol{M} \rangle \}$ |

with

$$\langle \boldsymbol{X}, \mathcal{L}_a \boldsymbol{X} \rangle = j \int_V \int_V \boldsymbol{X}_1 \cdot \boldsymbol{X}_2^* \, G_{12} \, \mathrm{dV}_1 \mathrm{dV}_2 \qquad \langle \boldsymbol{X}, \mathcal{L}_f \boldsymbol{X} \rangle = \frac{j}{k^2} \int_V \int_V \nabla_1 \cdot \boldsymbol{X}_1 \nabla_2 \cdot \boldsymbol{X}_2^* \, G_{12} \, \mathrm{dV}_1 \mathrm{dV}_2$$

$$\langle \boldsymbol{X}, \mathcal{L}_c \boldsymbol{X} \rangle = \frac{1}{2} \int_V \int_V \left( k\boldsymbol{X}_1 \cdot \boldsymbol{X}_2^* - \frac{1}{k} \nabla_1 \cdot \boldsymbol{X}_1 \nabla_2 \cdot \boldsymbol{X}_2^* \right) |\boldsymbol{r}_{12}| G_{12} \, \mathrm{dV}_1 \mathrm{dV}_2$$

$$\langle \boldsymbol{X}, \mathcal{K} \boldsymbol{Y} \rangle = \int_V \int_V \boldsymbol{X}_1^* \cdot [\boldsymbol{Y}_2 \times \nabla_1 G_{12}] \, \mathrm{dV}_1 \mathrm{dV}_2 = \int_V \int_V [\boldsymbol{X}_1^* \times \boldsymbol{Y}_2] \cdot \nabla_1 G_{12} \, \mathrm{dV}_1 \mathrm{dV}_2 = \int_V \int_V [\boldsymbol{Y}_1 \times \boldsymbol{X}_2^*] \cdot \nabla_1 G_{12} \, \mathrm{dV}_1 \mathrm{dV}_2$$

$$\langle \boldsymbol{X}, \mathcal{K}_c \boldsymbol{Y} \rangle = k^2 \int_V \int_V \boldsymbol{X}_1^* \cdot [\boldsymbol{Y}_2 \times \boldsymbol{r}_{12}] G_{12} \, \mathrm{dV}_1 \mathrm{dV}_2 = k^2 \int_V \int_V [\boldsymbol{Y}_1 \times \boldsymbol{X}_2^*] \cdot \boldsymbol{r}_{12} G_{12} \, \mathrm{dV}_1 \mathrm{dV}_2$$

expressions of Table I subject to the zero interior fields boundary condition (15). In a practical implementation, the domain of the optimization problem can be halved, so that the optimization runs with the electric current coefficients $\mathbf{I}$ as variables, while the magnetic current coefficients $\mathbf{M}$ are computed via (17) to ensure vanishing interior fields. Further constraints can be added, if a specific electric or magnetic dipole mode is desired. A good initial guess for $\mathbf{I}$ can be found using the characteristic mode approach, as described in the next Section.

## V. NUMERICAL EXAMPLES

In this section, we consider two canonical examples validating the proposed approach to determining the lower bound on $Q$ in general and the derived expressions for the energies and $Q$ in particular. To find the optimal currents and the minimum possible $Q$, the following simplified procedure is adopted

1) Find the optimal electric current ensuring minimum $Q_{\mathrm{J}}$ (air-core lower bound) using the characteristic mode approach [9]. Characteristic modes [31] are the solutions of a generalized eigenvalue problem formulated as

$$[\mathbf{X}]\mathbf{I} = \lambda[\mathbf{R}]\mathbf{I} \tag{18}$$

where $[\mathbf{X}] = \mathrm{Im}[\mathbf{Z}]$ and $[\mathbf{R}] = \mathrm{Re}[\mathbf{Z}]$. If the eigenvector $\mathbf{I}$ representing the electric current of each mode is normalized so that $\mathbf{I}^*[\mathbf{R}]\mathbf{I} = 1$, it is easy to show using (18) and Poynting's theorem that

$$\lambda = 4\omega(W^{\mathrm{m}} - W^{\mathrm{e}}) \tag{19}$$

implying that the lowest eigenvalue $\lambda$ corresponds to a mode closest to the resonance. By computing a few lowest eigenvalues and their respective eigenvectors, which appear to be mutually orthogonal [31], we can find the mode with the lowest $Q_{\mathrm{J}}$.

2) For the eigenvector $\mathbf{I}$ of the lowest $Q_{\mathrm{J}}$ mode, find the corresponding magnetic current cancelling the fields in the shape's volume using (17).
3) Compute the lower bound by applying the expressions in Table I or II to the superposition of the obtained electric and magnetic currents.

As a by-product, we also obtain the respective $Q$'s for electric and magnetic currents radiating in free space, $Q_{\mathrm{J}}$ and $Q_{\mathrm{M}}$.

At this point, it should be noted that to the best of the author's knowledge, no general proof showing that the characteristic modes of an arbitrarily shaped object actually contain the lowest $Q_{\mathrm{J}}$ mode has been reported so far. Nevertheless, as demonstrated by examples in [9] as well as here, the approach works well for simple shapes, such as spheroids, discs, cylinders, etc. Furthermore, the characteristic modes by Harrington and Mautz [31] allow us not only compute the lower bound on $Q$, but also investigate the mode structure of the shapes considered in this Section.

The operators in Table II are discretized using higher order method of moments [32] based on higher order curvilinear geometry representation and higher order hierarchical Legendre basis functions [33].

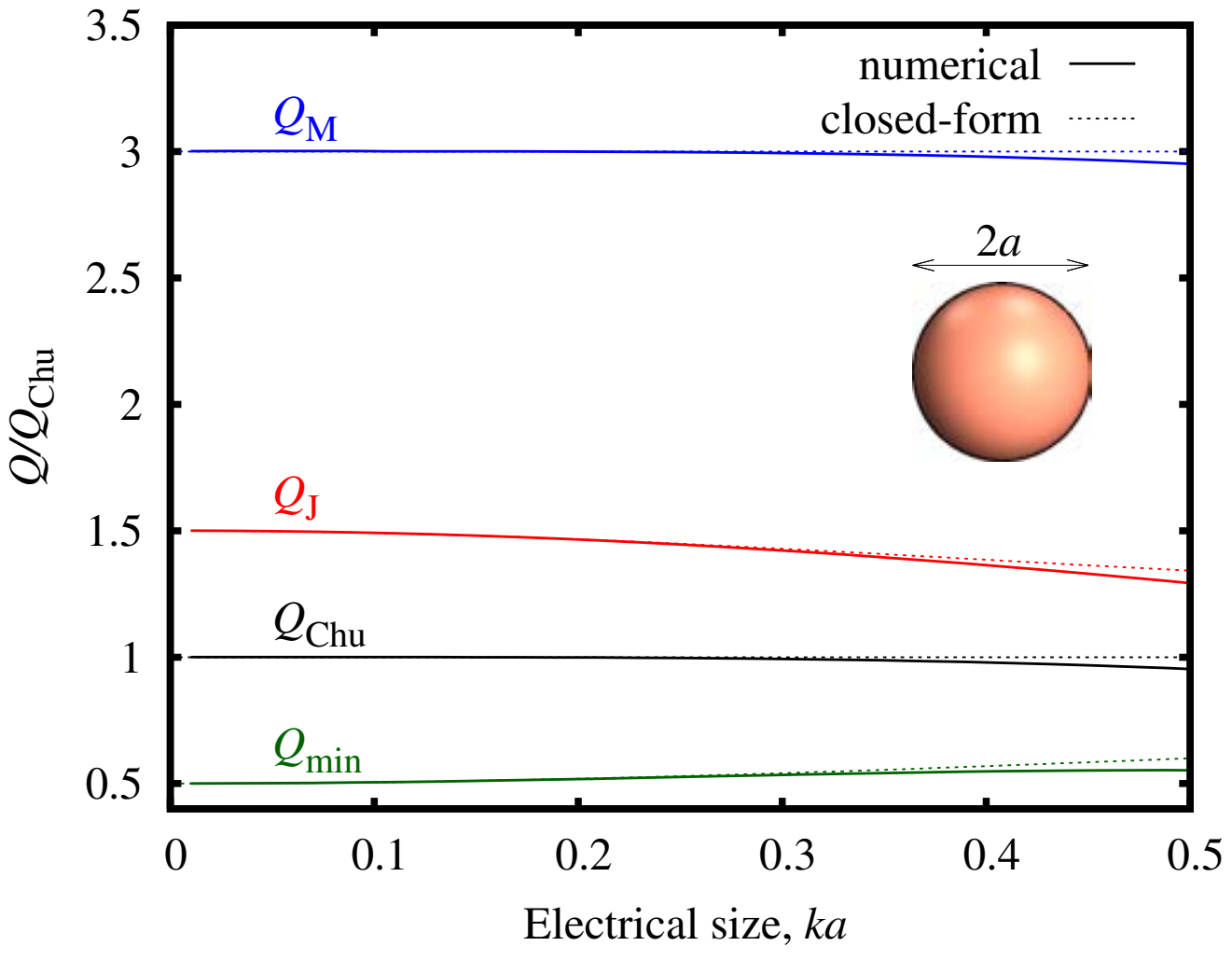

Fig. 1. Radiation $Q$ normalized to the Chu lower bound $Q_{\mathrm{Chu}}$ for an ideal spherical electric dipole antenna (radiating $\mathrm{TM}_{10}$ spherical mode) excited by electric current (labeled $Q_{\mathrm{J}}$), magnetic current ($Q_{\mathrm{M}}$), and a superposition of electric and magnetic currents ($Q_{\mathrm{Chu}}$). Also shows $Q$ for a dual-mode $\mathrm{TM}_{10}$-$\mathrm{TE}_{10}$ antenna excited by a superposition of electric and magnetic currents ($Q_{\mathrm{min}}$).

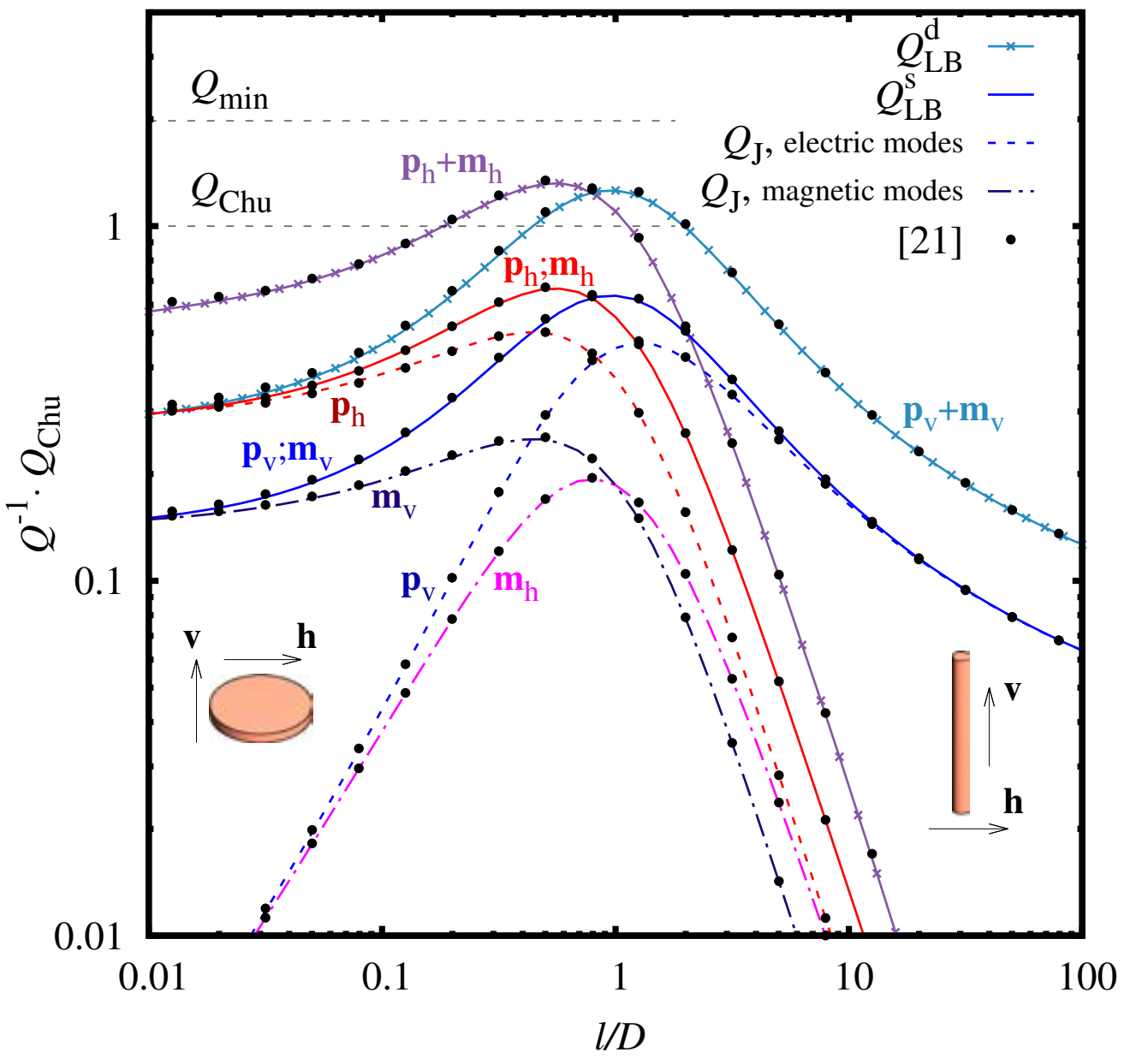

Fig. 2. Ratio of the Chu lower bound $Q_{\mathrm{Chu}}$ to the radiation $Q$ of electric and magnetic dipole modes excited by electric current in an ideal cylindrical antenna with $ka = 0.1$ versus its length to diameter ratio $l/D$. The respective dipole moments are denoted by **p** and **m**. Lower bounds for single- and dual-mode cylindrical antennas, $Q^{\mathrm{s}}_{\mathrm{LB}}$and $Q^{\mathrm{d}}_{\mathrm{LB}}$, are also presented.

### *A. Sphere*

First, we apply the above procedure to a spherical shape of radius $a$. The characteristic mode decomposition for a sphere recovers a set of spherical modes with the first three (corresponding to three lowest by magnitude eigenvalues) being the electric dipole modes ($\mathrm{TM}_{10}$ spherical mode) and the second three being the magnetic dipole modes ($\mathrm{TE}_{10}$ spherical mode). Each triplet consists of the respective degenerate dipole modes with mutually orthogonal moments.

For the electric current radiating one of the electric dipole modes, the magnetic current was determined via (17) and the $Q$'s corresponding to the electric current ($Q_{\mathrm{J}}$), the magnetic current($Q_{\mathrm{M}}$), and a combination of the electric and magnetic currents ($Q_{\mathrm{Chu}}$) were computed. The results are plotted in Fig. 1 (solid lines) as a function of the sphere electrical size $ka$, along with the curves representing respectively the closed-form bounds for the $\mathrm{TM}_{10}$ electric current (Wheeler-Thal bound [2], [15], [34]), for the $\mathrm{TM}_{10}$ magnetic current [3], and the Chu lower bound [1], [4] (dotted lines).

The two curves (obtained numerically and from the respective closed-form expression) in each of the three sets agree very well as $ka \to 0$, while the difference increases for larger electrical dimensions. This difference originates from the propagating energy density confined in the object volume (sphere, in this case). In the derivations in Section II, this density is subtracted from the total energy density, whereas in the closed-form bounds it is assumed to be zero. The latter assumption seems intuitively correct; however, nothing in the rigorous derivations of the $Q$ expressions (in Section II as well as in [12]) indicates that this part of the propagating energy should be excluded (see also [23], [24]). In any case, the difference is of the order of $ka$, and thus, of little practical importance.

The above exercise was repeated for the magnetic dipole mode; the resulting plot coincides with the one in Fig. 1, except that the curves for electric and magnetic currents swap – a natural consequence of the electromagnetic duality.

Finally, the electric and magnetic currents for an electric dipole mode and a magnetic dipole mode were normalized, so that these two modes radiate the same power; their superposition was then used to compute the $Q$. The result is shown in Fig. 1 under the label '$Q_{\mathrm{min}}$' denoting the minimum possible $Q$ for a linear passive time-invariant antenna [1], [20], [35]. Again, the curve is in perfect agreement (subject to the difference discussed just above) with the theoretical result, which reads as [22], [35]

$$Q_{\mathrm{min}} = \frac{1}{2}\left\{\frac{1}{(ka)^3} + \frac{2}{ka}\right\}. \tag{20}$$

### *B. Cylinder*

The six dipole modes, three electric and three magnetic, can also be excited in a cylinder, but only two modes in each triplet are degenerate – those with moments orthogonal to the cylinder axis. Similar to a sphere, dipole modes in a cylinder correspond to eigenvalues with the lowest magnitudes, an thus, are easily recovered in the characteristic mode decomposition. The sign of each eigenvalue indicates the type of the mode, according to (19) — negative for electric modes, and positive for magnetic modes. The corresponding eigenvector representing the electric current distribution for each mode is then inserted into the expressions in Table I or II to determine $Q_{\mathrm{J}}$. Figure 2 shows $Q_{\mathrm{J}}$ for all six dipole modes versus the length to diameter ratio ($l/D$) of the cylinder; in all cases, the size of the cylinder is such that it can be enclosed in a sphere of electrical radius $ka = 0.1$. The moments for electric and magnetic dipoles are denoted by $\mathbf{p}$ and $\mathbf{m}$, respectively; subscripts 'h' and 'v' correspond to dipole moments aligned perpendicular and along the cylinder axis.

When the energy stored inside the cylinder is excluded by applying step 2) of the procedure outline in the beginning of this Section, the resulting $Q$'s for $\mathbf{m}_\mathrm{v}$ and $\mathbf{p}_\mathrm{v}$ modes appear to be equal; in other words, the single-mode lower bound $Q_\mathrm{LB}^\mathrm{s}$ is the same for these two modes. Similarly, $\mathbf{m}_\mathrm{h}$ and $\mathbf{p}_\mathrm{h}$ modes have identical single-mode lower bounds.

We can also note in Fig. 2 that the $Q_\mathrm{LB}^\mathrm{s}$ bound approaches $Q_\mathrm{J}$ for $\mathbf{m}_\mathrm{v}$ and $\mathbf{p}_\mathrm{v}$ modes at the extremes of the $l/D$ ratio, that is, for a thin disk ($l/D \ll 1$), $Q_\mathrm{LB}^\mathrm{s}$ approaches $Q_\mathrm{J}$ for $\mathbf{m}_\mathrm{v}$, while for a thin dipole ($l/D \gg 1$) $Q_\mathrm{LB}^\mathrm{s}$ approaches $Q_\mathrm{J}$ for $\mathbf{p}_\mathrm{v}$.

Identical lower bounds for the modes in pairs $\mathbf{m}_\mathrm{h}$, $\mathbf{p}_\mathrm{h}$ and $\mathbf{m}_\mathrm{v}$, $\mathbf{p}_\mathrm{v}$ imply that they have equal, but opposite in kind, external stored energies for the same radiated power, and therefore, can be combined in a dual-mode self-resonant configuration. Dual-mode lower bounds $Q_\mathrm{LB}^\mathrm{d}$ for $\mathbf{m}_\mathrm{h} + \mathbf{p}_\mathrm{h}$ and $\mathbf{m}_\mathrm{v} + \mathbf{p}_\mathrm{v}$ mode pairs are shown in Fig. 2.

To validate the results, the $Q$'s presented in Fig. 2 were also calculated using approximate expressions from [21]. The expressions are based on static polarizabilities (were computed with COMSOL Multiphysics), and thus, valid in the case of vanishingly small antennas. In our case, the electrical size of the cylinder is $ka = 0.1$, which is small enough so the agreement with the results of this paper is very good.

There are optimum $l/D$ ratios that minimize the single- and dual-mode bounds; these are shown in Fig. 3a as functions of the electrical size $ka$ of an imaginary sphere enclosing the cylinder. The corresponding values of the minimum possible $Q$ are plotted in Fig. 3b. That fact that the single-mode bounds fall below the Chu lower bound for $ka > 0.9$ is again attributed to the uncertainty associated with the propagating energy density within the object. Accounting this energy in the Chu lower bound results in a bound ($Q_\mathrm{Chu} - ka$) [24], which is also shown in Fig. 3b.

Results of the quasi-static approach of [21] are also plotted in Fig. 3 for comparison. As expected, the agreement with the results of this paper is excellent in the limit of $ka \to 0$, while the deviation becomes appreciable at larger electrical sizes. Since the polarizabilities are independent of $ka$, the quasi-static approach yields constant optimum $l/D$ ratios. Furthermore, the optimum ratios in the single- and dual-mode cases are the same (for 'h-' or 'v-modes'), because the single- and dual-mode $Q$'s are simply related by a factor of 2 in the limit of $ka \to 0$. In the case of finite $ka$, the behavior, however, is more complicated manifesting itself in splitting of the $l/D$-curves. It is observed in Fig. 3b that the $Q$ bounds predicted by the quasi-static approach are rather optimistic; at $ka = 0.5$, the values are 12% and 24% below the $Q$'s of this paper in the single- and dual-mode case, respectively.

## VI. Conclusions

A technique for determining the lower bounds on $Q$ for finite size antennas of arbitrary shape is presented. The technique combines electric and magnetic current densities on an exterior surface of an antenna shape in such a way that the fields interior to the surface are zero, which implies that the stored energy in the antenna volume is excluded from the total stored energy, and thus, $Q$. The latter is then the true lower bound for the given shape.

The computation of $Q$ is based on new expressions for the stored energy, radiated power, and $Q$ of coupled electric and magnetic currents in free space. These expressions can be incorporated into a method of moment code in a straightforward manner, enabling not only computation of the lower bounds on $Q$ but also optimization for minimum $Q$ (and thus, for maximum frequency bandwidth) of microstrip antennas as well as antennas loaded with magneto-dielectric material. In practice, the inverse of $Q$ can be used to estimate the impedance bandwidth of single-resonance antennas up to about 0.3 wavelengths in size ($ka \approx 1$), or as long as $Q \gtrsim 10ka$.

In contrast, the lower bound on $Q$ by the quasi-static ap-proach of [21] is accurate only for vanishingly small antennas ($ka \ll 1$), as illustrated in Fig. 3. Besides that, the quasi-static approach requires a separate static solver to numerically find polarizabilities for shapes more complex than simple spheroids.

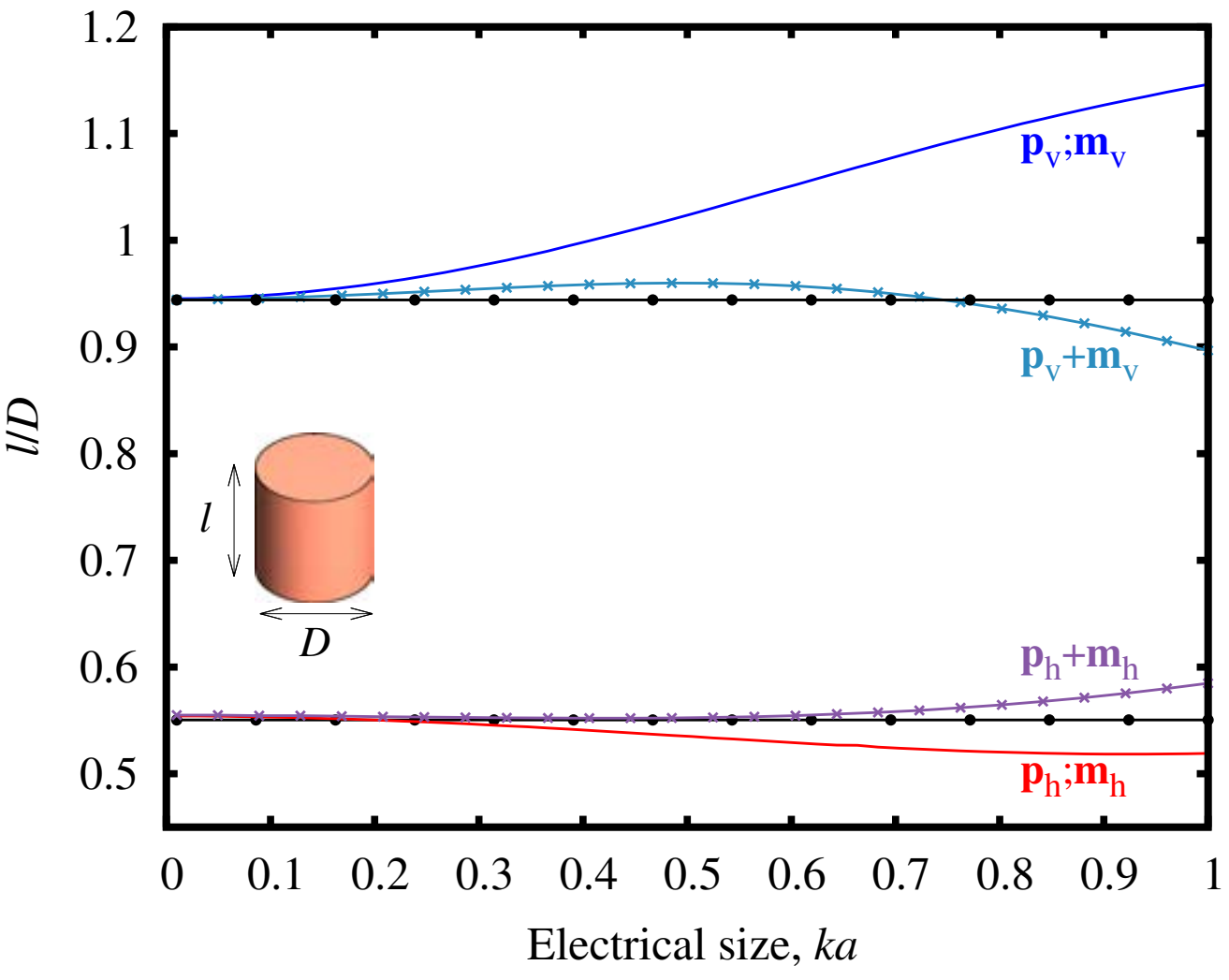

(a)

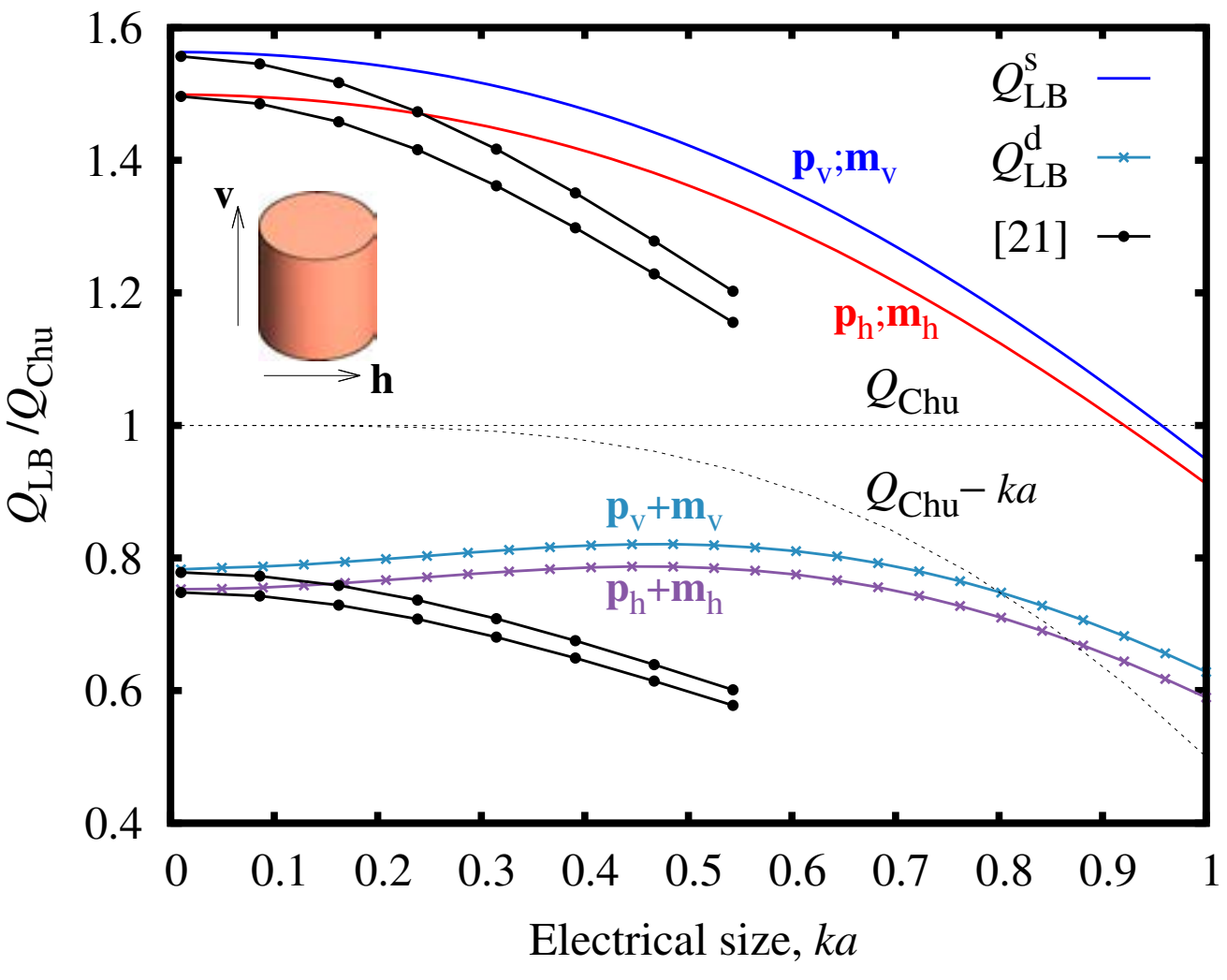

(b)

Fig. 3. Optimum $l/D$ ratio (a) and the corresponding minimum possible $Q$ (b) for single- and dual-mode configurations versus electrical size $ka$ of a sphere enclosing the cylinder.

Numerical examples provided in Section V show that a simplified procedure for finding the lower bound on $Q$ outlined at the beginning of that section works for finite size spheres and cylinders and it was also verified for spheroids. This indicates that it may work for other generic shapes as well. For more complex geometries, the simplified procedure can be used as a good initial guess for an optimization aiming at the minimum $Q$, as suggested in Section IV.

## Appendix A

Insert (4) into the last term in (6) and integrate over the entire space $V_\infty$ as

$$\begin{aligned}
&\int_{V_\infty} \nabla\Phi^* \cdot [\nabla \times \boldsymbol{F}]\mathrm{dV} \\
&= \frac{j}{\omega}\int_{V_\infty} \left\{ \int_V (\nabla \cdot \boldsymbol{J}_2^*)\, \nabla G_2^* \mathrm{dV}_2 \cdot \int_V [\boldsymbol{M}_1 \times \nabla G_1]\mathrm{dV}_1 \right\} \mathrm{dV} \\
&= \frac{j}{\omega}\int_V \int_V (\nabla \cdot \boldsymbol{J}_2^*)\boldsymbol{M}_1 \int_{V_\infty} [\nabla G_1 \times \nabla G_2^*]\mathrm{dV}\,\mathrm{dV}_1\,\mathrm{dV}_2. \quad (21)
\end{aligned}$$

The inner integral in the last line of (21) was treated in [12, equation (108)], where it was shown that

$$\begin{aligned}
&\int_{V_\infty} [\nabla G_1 \times \nabla G_2^*]\mathrm{dV} \\
&= \int_{V_\infty} \nabla \times (G_1 \nabla G_2^*) - G_1(\nabla \times \nabla G_2^*)\mathrm{dV} \\
&= \lim_{S_\infty \to \infty} \oint_{S_\infty} G_1[\hat{\boldsymbol{r}} \times \nabla G_2^*]\mathrm{dS} \quad (22)
\end{aligned}$$

where $S_\infty$ is a surface enclosing the volume $V_\infty$. The surface integral in (22) is zero, because $\nabla G_2^*$ and $\hat{\boldsymbol{r}}$ are collinear for $S_\infty \to \infty$.

## Appendix B

Take the gradient with respect to $r_1$ of both sides of identity (64) in [23] as

$$\begin{aligned}
&\nabla_1 \int_{V_\infty} G_1 G_2^* - \frac{e^{jk(\boldsymbol{r}_1-\boldsymbol{r}_2)\cdot\hat{\boldsymbol{r}}}}{16\pi^2|\boldsymbol{r}|^2}\mathrm{dV} \\
&= \nabla_1 \left[ -\frac{\sin(k|\boldsymbol{r}_{12}|)}{8\pi k} \right. \\
&\left. -j\frac{|\boldsymbol{r}_1|^2 - |\boldsymbol{r}_2|^2}{8\pi k^2}\left(\frac{\sin(k|\boldsymbol{r}_{12}|)}{|\boldsymbol{r}_{12}|^3} - \frac{k\cos(k|\boldsymbol{r}_{12}|)}{|\boldsymbol{r}_{12}|^2}\right)\right]. \quad (23)
\end{aligned}$$

Bringing the gradient inside the integral and recognizing that $\nabla_1 G_1 = -\nabla G_1$ we get

$$\begin{aligned}
&\int_{V_\infty} -(\nabla G_1)G_2^* - \hat{\boldsymbol{r}}jk\frac{e^{jk(\boldsymbol{r}_1-\boldsymbol{r}_2)\cdot\hat{\boldsymbol{r}}}}{16\pi^2|\boldsymbol{r}|^2}\mathrm{dV} \\
&= -\frac{\boldsymbol{r}_{12}}{|\boldsymbol{r}_{12}|}\frac{\cos(k|\boldsymbol{r}_{12}|)}{8\pi} \\
&\quad -j\frac{2\boldsymbol{r}_1}{8\pi k^2}\left(\frac{\sin(k|\boldsymbol{r}_{12}|)}{|\boldsymbol{r}_{12}|^3} - \frac{k\cos(k|\boldsymbol{r}_{12}|)}{|\boldsymbol{r}_{12}|^2}\right) \\
&\quad -j\frac{\boldsymbol{r}_1 + \boldsymbol{r}_2}{8\pi k^2}\left(\frac{k^2\sin(k|\boldsymbol{r}_{12}|)}{|\boldsymbol{r}_{12}|}\right. \\
&\quad \left. - 3\left(\frac{\sin(k|\boldsymbol{r}_{12}|)}{|\boldsymbol{r}_{12}|^3} - \frac{k\cos(k|\boldsymbol{r}_{12}|)}{|\boldsymbol{r}_{12}|^2}\right)\right). \quad (24)
\end{aligned}$$

Substituting $\boldsymbol{r}_1 = 0.5(\boldsymbol{r}_1 - \boldsymbol{r}_2) + 0.5(\boldsymbol{r}_1 + \boldsymbol{r}_2)$ and multiplying (24) with $-j$, we finally obtain the identity

$$\begin{aligned}
&\int_{V_\infty} j(\nabla G_1)G_2^* - \hat{\boldsymbol{r}}k\frac{e^{jk(\boldsymbol{r}_1-\boldsymbol{r}_2)\cdot\hat{\boldsymbol{r}}}}{16\pi^2|\boldsymbol{r}|^2}\mathrm{dV} \\
&= j\frac{\boldsymbol{r}_{12}}{|\boldsymbol{r}_{12}|}\frac{\cos(k|\boldsymbol{r}_{12}|)}{8\pi} \\
&\quad -\frac{\boldsymbol{r}_{12}}{8\pi k^2}\left(\frac{\sin(k|\boldsymbol{r}_{12}|)}{|\boldsymbol{r}_{12}|^3} - \frac{k\cos(k|\boldsymbol{r}_{12}|)}{|\boldsymbol{r}_{12}|^2}\right) \\
&\quad -\frac{\boldsymbol{r}_1 + \boldsymbol{r}_2}{8\pi k^2}\left(\frac{k^2\sin(k|\boldsymbol{r}_{12}|)}{|\boldsymbol{r}_{12}|}\right. \\
&\quad \left. - 2\left(\frac{\sin(k|\boldsymbol{r}_{12}|)}{|\boldsymbol{r}_{12}|^3} - \frac{k\cos(k|\boldsymbol{r}_{12}|)}{|\boldsymbol{r}_{12}|^2}\right)\right) \quad (25a) \\
&= j\frac{\boldsymbol{r}_{12}}{2}\mathrm{Re}\{G_{12}\} - \frac{1}{2k^2}\mathrm{Im}\{\nabla_1 G_{12}\} \\
&\quad + \frac{\boldsymbol{r}_1 + \boldsymbol{r}_2}{2k^2}\mathrm{Im}\{k^2 G_{12} + \frac{2\boldsymbol{r}_{12}}{|\boldsymbol{r}_{12}|^2}\nabla_1 G_{12}\}. \quad (25b)
\end{aligned}$$

The last term in (25a) and (25b) depends on the choice of the coordinate system, which, along with the coordinate dependence of the stored energies associated with the electric (and magnetic) current [23], makes the total stored energy, and thus $Q$, coordinate dependent as well. This controversial aspect was first noted in [24]. Nevertheless, as argued in [12], the coordinate independence can formally be restored by setting $\boldsymbol{r}_1 + \boldsymbol{r}_2 = 0$, which is equivalent to placing the origin of the coordinate system in the middle between the points $\boldsymbol{r}_1$ and $\boldsymbol{r}_2$.

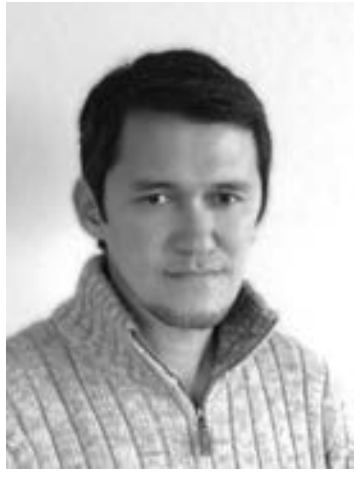

**Oleksiy S. Kim** received the M.Sc. and Ph.D. degrees from the National Technical University of Ukraine, Kiev, in 1996 and 2000, respectively, both in electrical engineering. In 2000, he joined the Antenna and Electromagnetics Group at the Technical University of Denmark (DTU). He is currently an associate professor with the Department of Electrical Engineering, Electromagnetic Systems, DTU.

His research interests include electrically small antennas, computational electromagnetics, metamaterials, photonic bandgap and plasmonic structures.